# wHealth - Transforming Telehealth Services

Rajib Rana[^,*], Margee Hume[+], John Reilly[#], Jeffrey Soar[^]

[^]University of Southern Queensland, Australia; [+]University of central Queensland, Australia; [#]Townsville Hospital Mental Health Service Group

## Abstract

*A worldwide increase in proportions of older people in the population poses the challenge of managing their increasing healthcare needs within limited resources. To achieve this many countries are interested in adopting telehealth technology. Several shortcomings of state-of-the-art telehealth technology constrain widespread adoption of telehealth services. We present an ensemble-sensing framework - wHealth (short form of wireless health) for effective delivery of telehealth services. It extracts personal health information using sensors embedded in everyday devices and allows effective and seamless communication between patients and clinicians. Due to the non-stigmatizing design, ease of maintenance, simplistic interaction and seamless intervention, our wHealth platform has the potential to enable widespread adoption of telehealth services for managing elderly healthcare. We discuss the key barriers and potential solutions to make the wHealth platform a reality.*

## Introduction

The world's population is ageing, with the proportion of older people gradually rising. In Australia the population aged 75 or more years is expected to rise by 4 million from 2012 to 2060, increasing from about 6.4 to 14.4 per cent of the population (Productivity Commission, 2013; Turner & McGee-Lennon, 2013). In the UK the old-age dependency ratio, which represents the population aged 65+

---

[*] **Corresponding Author**

Rajib Rana, Department of Management and Enterprise, USQ Springfield campus, Springfield Central Qld 4300, Australia

(Email: to.rajib.rana@gmail.com)

as a proportion of those aged 16-64, was 24.4% in 2000 and is expected to become 39.2% by 2050 (London: Authority of the House of Lord, 2003). In Europe, the number of older people is expected to grow from 75 million in 2004 to 133 million in 2050 (Bovenberg & Van der Linden, 1997). Japan will perhaps face the greatest increase with the old-age dependency ratio projected to be 71.3% by 2050 (London: Authority of the House of Lord, 2003).

This substantial increase in the elderly population poses significant challenges in provision of health care. An overview of improving health services for older people suggests that the key components are supporting independence in healthy active ageing, living well with stable long term conditions and complex co-morbidities, rapid support close to home in crisis, good acute hospital care when needed, good discharge planning and post-discharge support, rehabilitation after acute illness, high quality residential care for those in need supported by choice, control and support towards the end of life and integration to provide person-centred co-ordinated care (Oliver et al., 2014). Although the current evidence for effectiveness of telehealth is limited(Rabasca, 2000), it is recognised that telehealth has the potential to enhance service delivery in all components of community based care, within a context of integrated locality based services(Frueh, Henderson, & Myrick, 2005). There are several challenges to enhancing the effectiveness of the telehealth technology. Devices used for telehealth can pose barriers to use due to their size and poor design (Kang et al., 2010). In many instances a large number of devices are needed for one function. Carranza et al. (2010) reported the use of six different sensors just to detect falls. Many of these devices are not customisable with not user-friendly interfaces (Rahimpour, Lovell, Celler, & McCormick, 2008). We propose an ensemble-sensing framework – wHealth that through using only everyday devices provides non-stigmatising and seamless monitoring of various physiological, physical and behavioural parameters and enables effective communication between patients and clinicians. We believe the proposed wHealth platform will overcome the barriers of telehealth services enabling its widespread adoption for managing elderly healthcare.

**Barriers for wide-adoption of Telehealth Technology**

*Limited opportunity for passive monitoring*

Passive monitoring has the inherent potential benefit of obviating the problems associated with incorrect use and subject compliance (Bowes, Dawson, & Bell, 2012). It also provides effective care coordination tools, which support the professional caregivers' efficiency and reductions in workloads, as well as significant reductions of billable interventions and hospital days and hence cost of care to payers. It may be one of the key solutions to the problem of care delivery to the world's growing elder population, and has potential to provide health care systems, as well as social services with capacity to extend delivery

(Alwan et al., 2007) of health and social services. Despite the need, only a few studies have involved passive monitoring (Bensink, Hailey, & Wootton, 2007), primarily due to the lack of availability of easily wearable physiological monitors that can seamlessly and reliably record or transfer health data.

*Need for Managing Multiple Devices*

Existing home telehealth systems often comprise multiple physiological monitors (e.g., blood pressure or heart rate monitor, thermometer etc.). The user needs to ensure that all have sufficient power and are properly calibrated. This is particularly difficult for elderly due to increasing impairment and disability and/or unfamiliarity with technology. Telehealth technology requiring many devices is thus at risk of low adherence. For wider acceptability and usability multiple functionalities need to be simplified and incorporated in one device where possible.

*Electromagnetic Incompatibility*

Electromagnetic compatibility (EMC) is the ability of electronic devices to co-exist without adversely affecting each other's performance (Yu, Wu, Yu, & Xiao, 2006). Various physiological monitoring devices in the telehealth platform could be manufactured by multiple manufacturers. Most manufacturers use proprietary firmware and standard, which can potentially cause electromagnetic incompatibility, with potential malfunctioning. If multiple monitors (such as ECG, physical activity, skin conductance sensor on wrist band) are embedded in one device, this problem can be alleviated or minimized.

*Limited Focus on Device Design*

Telehealth devices often need clear displays and larger keyboards for people with declining visual acuity. It is important that they avoid "hospital-grade" device appearance to avoid rejection by users. The best trade-off between functionality and design requires work on branding and consumer awareness and acceptability rather than considering telehealth as a functional necessity.

*Functionality and interaction*

Telehealth technologies need to be acceptable to a wide variety of age ranges, expertise, capabilities, and preferences. This will require enhanced functionality as users become more familiar with the systems and user interfaces that will need to flexibly adapt to specific needs. Modern telehealth systems can also provide an interactive interface, which works as both data collection hub and a visualization medium for personal health data. Visualisation of personal health data on these devices often needs complex operation (Kidholm, Dyrvig, Dinesen, & Schnack), resulting in limited or no benefit to its user. Wider adoption of telehealth technology requires that these devices should be intuitive and simple, with readily accessible functionalities.

*Lack of Customisation*

Many existing telehealth solutions are not easily customisable. Telehealth technology should be flexibly tailored to individual preferences (Clark & McGee-Lennon, 2011), thereby making it more accessible and able to fit more seamlessly into a person's lifestyle. Smartphones/tablets can facilitate telehealth services through increasing ease of application level customisation to suit individuals.

*Lack of Acceptance*

A challenge in the adoption of Telehealth technology has been the perceived stigmatisation of users (Turner & McGee-Lennon, 2013), Wearing an emergency alarm for example can be seen by users as a sign of ageing and loss of independence (Turner & McGee-Lennon, 2013). Fear of technology (Smith & Maeder, 2010) and concerns about higher direct costs (Cimperman, Brenčič, Trkman, & Stanonik, 2013) of a telehealth systems might also be a reason for low adoption. Acceptance levels may vary with social context and the consumer's technological sophistication and attitude. A non-stigmatising design of monitoring devices, user-friendly and customisable interface, maximum control over data sharing and privacy, can be expected to increase acceptance of telehealth services.

*Higher Costs*

Perceptions about cost is one of the most critical factors restricting the widespread adoption of telehealth technology. Purchasing and maintaining monitoring devices often incur higher cost. If physiological monitors can be built into day-to-day products it can be expected these will deliver telehealth functionality at lower cost. In the UK, also in most EEC[†] countries users need to buy telehealth devices using personal budget (Turner & McGee-Lennon, 2013). Reductions in the acquisition and operating costs of technologies and support services can be expected to encourage participation.

**Ensemble Sensing Framework**

"Ensemble" refers to a group of items viewed as a whole rather than individually. Our proposed ensemble-sensing framework – wHealth, is presented in Figure 5, wherein the three components: smartphone, smart watch and smart glass work in ensemble. Plethora of sensors embedded in these components allows profiling various behavioural, physical and physiological conditions of its user. Due to flexible programming options (on all three components), this framework offers customised passive sensing that can be tailored to meet individual's requirements. Within the wHealth framework the components create a symbiosis by sharing processing capacity, battery life and sensing opportunity. Smartphones have higher computational capacity, but limited sensing opportunity, as phones are not carried for extended periods. Whereas, smart watches and smart glasses are worn for relatively longer periods, but lack computational and battery power. In the wHealth framework, smart watch and smart glass transfer data to the smartphone for further processing and thus make the best use of battery life, computational power and sensing opportunity.

---

[†] European Economic Countries

*Smart Watch*

Unlike their predecessors, current smart watches are promoted as a fashion Accessory. Elegant design, durability and small form factor may make smart watches more desirable. The latest smart watches embed a plethora of sensors enabling monitoring. The block diagram of a smart watch in Figure 2 shows various embedded sensors such as 3-axis accelerometer, photoplethysmograph (heart rate variability, stress, relaxation), temperature and heat flux (activity and context) and electrodermal activity monitor (arousal and excitement by skin conductance). For communicating with smartphone or computer, smart watches use low-power communication protocols such as Bluetooth 4.0 and Near Field Communication (NFC). Smart watches also have relatively large displays, which are touch-operated.

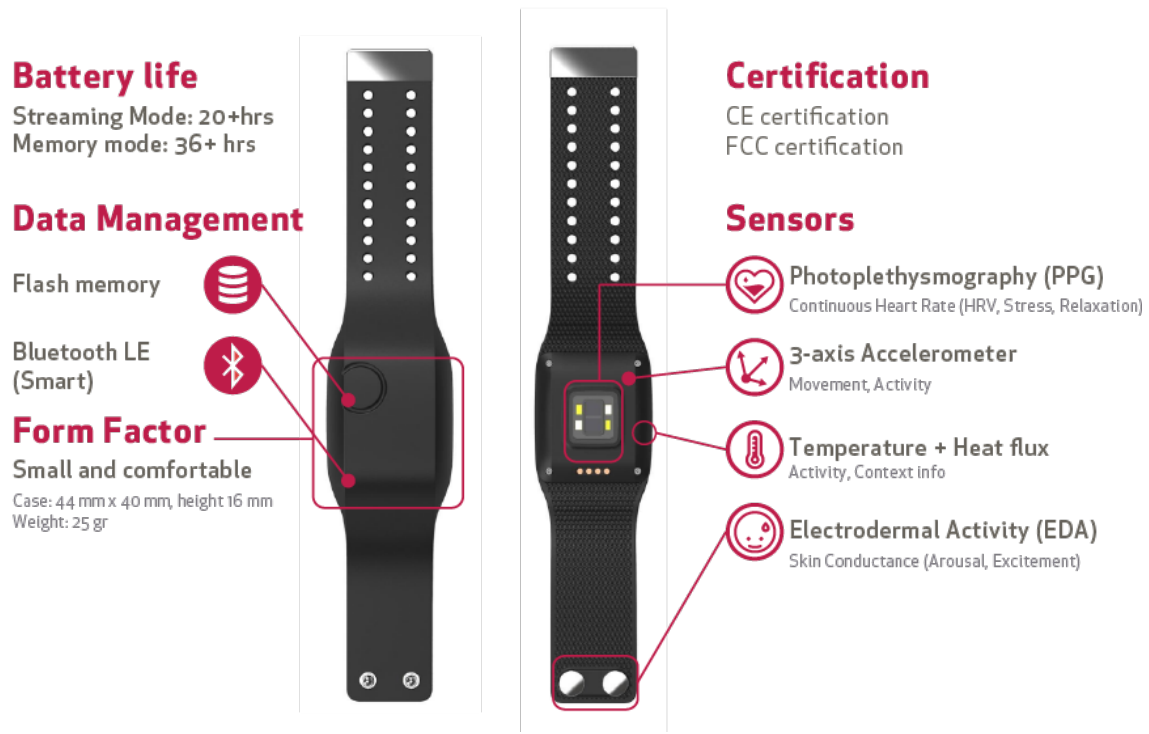

**Figure 2.** Smart watch sensors. [Source: internetmedicine.com]

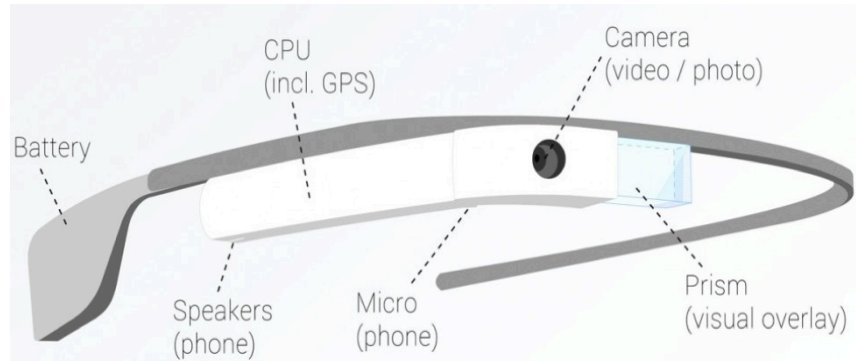

**Figure 3.** Smart glass sensors. [Source: TechNorms]

*Smart Glass*

Smart Glass is a wearable technology with an optical head-mounted display, which can either be attached to any spectacles or built into spectacles. Using smart glass wearers can communicate with Internet using voice commands. Of the many currently available glasses, Google glass (Google) is the most popular (Klonoff, 2014). It is highly fashionable and extremely lightweight.

A block diagram of smart glass is shown in Figure 3. Various sensors present on Google glass include camera, 3-axis gyroscope, 3-axis accelerometer, 3-axis magnetometer (compass), ambient light sensor, proximity sensor and bone conduction audio transducer. Google glass also has a prism like screen with a reflective surface. Images from Google Glass project onto this reflective surface, which redirects the lights towards wearer's eyes. The images are semi-transparent; it is possible to see through them to the real world on the other side.

*Smartphone*

Smartphones integrate mobile phone capabilities with features of a handheld computer or PDA. Their advanced sensing and communication capacity has been supplemented by an increased focus on optimising battery life, providing fast Internet connectivity and developing large and sensitive display. Smartphones allow numerous functionalities spanning managing finance to controlling home automation.

A block diagram (illustrating various on-board sensors) of smartphone in Figure 4 shows 3-axis accelerometer, gyroscope, magnetometer (compass), pressure sensor, front and rear camera, proximity sensor and temperature and humidity sensor. An interesting aspect of smartphone is its capacity to simultaneously connect to many external devices/monitors using standard communication protocol such as Bluetooth and NFC. It provides numerous Application Programming Interfaces (APIs) to flexibly configure various components of the phone. An overwhelming number of applications are also readily available to run on smartphones.

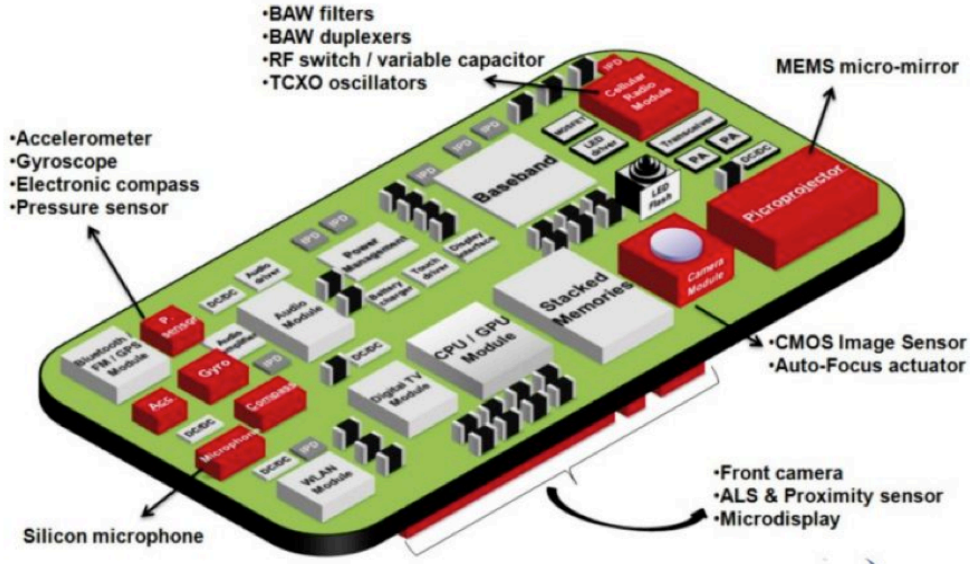

**Figure 4.** Smartphone sensors. [Source: Yole Development]

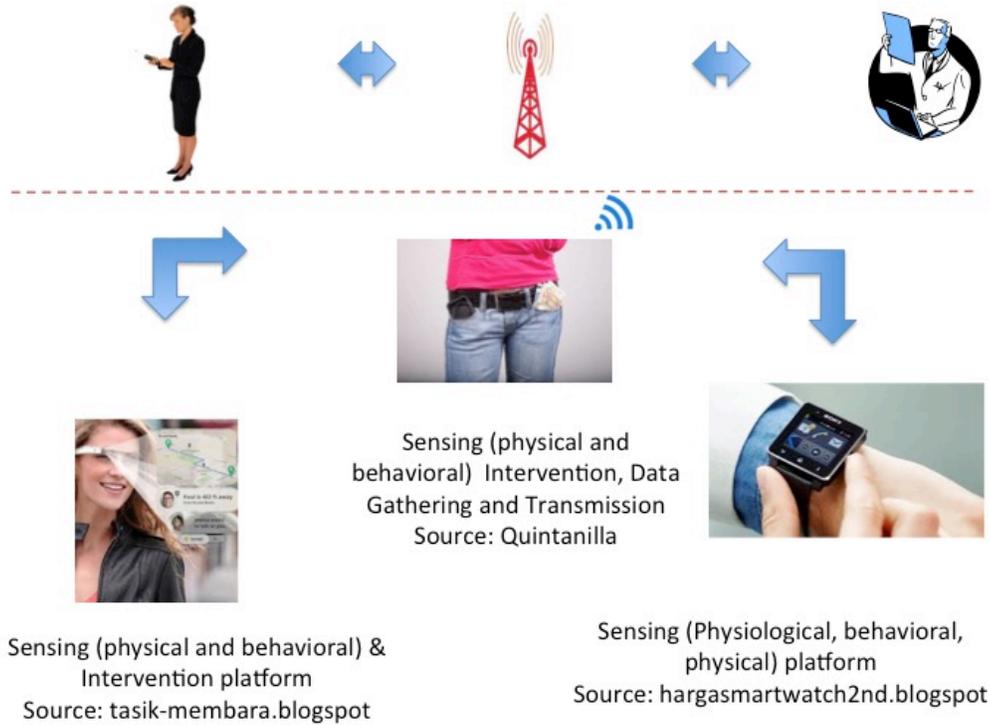

**Figure 5.** wHealth Framework.

## *Functionalities of the Ensemble Sensing Framework*

### *Behavioural sensing*

Behavioural sensing refers to sensing of sleep patterns, feeding and social interaction. The 3-axis accelerometer, magnetometer and gyroscope can be used to determine sleep patterns, physical activities and feeding. All three wHealth components have these sensors. Sensor data from three different locations in the body will help in better (compared to sensor data from one location) characterising various activities underlying these behaviours. Call log, GPS traces, Wi-Fi footprint, and Bluetooth trace will provide rich information about the number of calls made, places and person visited, which can be used to characterise social interactions.

### *Physiological sensing*

Seamless and passive physiological sensing has become possible with the rapid advancement of the smart watch technology. The smart watch is always in touch with the skin thereby providing opportunity to measure many physiological parameters, including heart rate variability, body temperature, skin conductance and blood pressure. Smartphones also allow sensing of affect using front camera and microphone, which can be used to potentially predict the onset of depressed episodes. Such information can be gathered opportunistically, such as when people use the phone for sending text or making phone call.

### *Physical sensing*

Physical sensing will profile the physical activeness of the user. The accelerometer sensor on the smart watch can be used to determine the ratio between active versus inactive periods. Specific activities such as walking, running, climbing upstairs, sitting and lying can also be determined using accelerometer data from the smart watch. Due to limited processing capacity and battery life, the above activity classification cannot be completely done on the watch. Summary data can be wireless transferred from the watch to the phone, which can then be used for classification.

### *Facilitating Interventions*

The smart glass can be utilized for providing seamless intervention. Using smart glass information can be projected just in front of the eye, which obviates the necessity to carry an additional device to receive instructions. Smartphones can also play an important role in enabling remote interventions. Cognitive behavioural therapy or brain games can be downloaded on patient's phone as a part of interventions. A phone can also be used as a platform for video and audio conferencing with clinicians, maintaining weight logs, activity diary and mood diary.

*Data collection and dissemination*

In the proposed wHealth framework, an external data collection and dissemination hub is not necessary. All data can be mediated on the smartphone and later on 3G/4G or Wi-Fi communication can be used to transfer the data to cloud storage for further processing. The standard, fast, low-power and non-proprietary communication protocol - Bluetooth 4.0, can be used to interconnect the devices; therefore it is possible to wirelessly connect the components almost instantly.

**Benefits of W-health platform**

*Passive Sensing*

The wHealth platform enables continuous collection of physical, physiological, behavioural data without any user intervention, allowing the detection of time-critical events such as falls. It can also help identify slowly developing conditions such as cognitive decline due to neurodegenerative disorders enabling earlier prevention measures. Facilitating passive sensing wHealth platform can reduce the burden of constantly monitoring health for both patient and carer.

*Non-stigmatising*

A non-stigmatising platform is the key focus of the wHealth framework. The components/devices in our framework are symbols of fashion and do not reveal themselves as monitoring devices; instead their extended functionalities can attract people of all ages.

*Ease of Maintenance*

The benefit of piggybacking sensing on everyday devices is that people will maintain them for their day-to-day functions. People are likely to recharge smart watch to tell the time. Similarly, they would recharge smartphone to maintain connections with friends and family. People will also recharge the glasses to see meeting reminders/emails in front of their eyes just in time. In this framework, physiological sensors are all in one device (smart watch), substantially reducing the number of devices and hence the maintenance overheads.

*Compatibility*

All three components of this framework can run the same Android Operating system, obviating the compatibility issues in some other market solutions. Most smart devices use standard Bluetooth as a communication protocol. This unified communication protocol allows these components to connect with each other almost instantly.

*Higher Acceptability*

Acceptability of any telehealth monitoring device depends on two key factors (Turner & McGee-Lennon, 2013). First, whether the device allows sufficient control over the data production and sharing and second, aesthetics. Due to open and non-proprietary development platform, full control is achievable on all three components of the wHealth platform. The user can fully control how the data is produced and to whom and to what extent it is shared. Access control is configurable; the user can adapt the data production and sharing if his/her preference changes over time.

From an aesthetics point of view, smart glasses (such as Google glass) are quite lightweight and fashionable, and they fit very nicely with most of the glass frames. Google has partnered with the Italian eyewear company Luxottica, owners of Ray-Ban, Oakley and other brands to offer additional frame designs. Similarly, the Apple smart watch blends fashion and technology elements into one beautiful, harmonious and amazingly crafted device. The modern and elegant design and numerous functionalities of both smart watch and smart glass have already created desire. The attractive design of these two components will create a non-stigmatising platform to reliably gather personal health data.

*Highly Customisable*

Both smart glass and smart watch are customisable in hardware and software. Watch straps can be changed to adapt colour and glass frames can be changed to adapt to new glasses designs. Many designer glasses are already compatible with Google glass attachment.

Open source software is used in both glass and watch, giving greater control on their functionality. Application developers can write applications in collaboration with consumers, so that the final product can fully meet the end-user requirements. Applications can also be developed with high configurability so that it can meet the requirement of any age group.

*Affordable*

Smartphones are available for as little as A$100. A latest smart watch can be purchased at around A$250-A$300. Google glasses are relatively expensive in the current market, however with technology advancing rapidly and increasing competition, soon smart glasses will be cheaper. The components of our framework also have their usual functions (such as, phones are used to keep connected with peer etc.), the cost is dispersed to achieve sensing functionalities as well as day-to-day functions.

**Challenges of W-Health platform**

There are two key challenges inherent to the W-Health platform, which need to be addressed to make it a reality.

*Privacy*

Privacy has been found to be a significant concern for participants of major Smartphone sensing studies (Dennison, Morrison, Conway, & Yardley, 2013; Seko, Kidd, Wiljer, & McKenzie, 2014). W-Health data that can be exploited to infer private information include (Christin, Reinhardt, Kanhere, & Hollick, 2011) (1) time and location, (2) sound samples, (3) pictures and videos (4) acceleration and (5) physiological data. Analysis of the frequency of visits to hospital may reveal someone's medical condition (Christin et al., 2011). By analysing characteristic sound patterns that are unique to certain events, location and presence in that event can be determined (Christin et al., 2011). Pictures or videos containing points of interest can reveal a user's location (Christin et al., 2011). If the mobile phone is carried on the body, information about gait and a user's identity may be inferred (Derawi, Nickel, Bours, & Busch, 2010). Lastly, intercepted medical information can be used by medical product sales companies, and health insurance companies to advertised related products.

Privacy overlays proposed (Parate, Chiu, Ganesan, & Marlin, 2013) can be used to identify the above reconstruction type attacks. Other privacy preserving measures suitable for the W-Health platform include (Christin et al., 2011) (1) pseudonymity, (2) spatial cloaking, and (3) data perturbation. Pseudonymity suggests all interaction with the application is performed under an alias user name. To achieve spatial cloaking k-anonymity (Sweeney, 2002) can be used where a group of k-participants are assigned same attribute (suburb name instead of exact location). Data perturbation perturbs the sensor samples by adding artificial noise (e.g., Gaussian noise) to the sensor data to determine community trends and distributions without revealing individual data (Christin et al., 2011).

In order to ensure data security many applications process information on-phone or locally without transferring information to cloud server, e.g., BodyBeat (Rahman et al., 2014) locally processes audio data between a wearable sensor and a smartphone. MoodScope (LiKamWa, Liu, Lane, & Zhong, 2013) also processes data locally. Resource incentive tasks such as predictive model construction are still conducted at resourceful servers, where privacy invasion can occur (LiKamWa et al., 2013). Techniques need to be developed to train predictive models using data from multiple people without invading individual's privacy (LiKamWa et al., 2013). Furthermore, due to the high prevalence of smartphones with multiple sensors, actuators, processors and wireless communication capabilities, privacy and information security issues have also started to emerge from the physical environment of a user (Konings & Schaub,

2011). These include gathered information from sensors (observers) in the user's proximity or by actuators (distributors) disturbing the user with intervening actions. To maintain data security in these challenging physical environments "Territorial" privacy (Konings & Schaub, 2011) can be utilized that allows users to control and monitor their privacy in the presence of observers and disturbers while moving between private (home) and public (office) environments.

*Limited Battery Power*

Despite advancements in the smartphone technology limited battery life is still a major constraint on the smartphone platform (R. Rana, Chou, Bulusu, Kanhere, & Hu, 2015; R. K. Rana, Chou, Kanhere, Bulusu, & Hu, 2010; Voida et al., 2013). Study participants from the mobile phone intervention studies have identified limited smartphone battery life as one of the most common problems (Burns et al., 2011; Dennison et al., 2013). Similarly, limited battery power is a major constrain for both smart glass (Muensterer, Lacher, Zoeller, Bronstein, & Kübler, 2014) and smart watch (Mooring & Fitzgerald, 2012). Using cloud-offloading, smartphone processing tasks can be delegated to a resourceful server; however this is only beneficial when the computational requirement is much higher than that of transmission (Kumar & Lu, 2010). The wHealth platform would generate enormous amount of data, which will take-up much transmission causing the battery to deplete rather quickly.

Low-power co-processors are becoming available on smartphones, which need to be utilized to achieve power savings (and thus extend battery life) by minimizing involvement of the power-hungry CPU. Readily available example of co-processors is the Natural Language Processor (NLP) and Contextual Computing Processor (CCP) available on the Motorola X8 system. Lane et al. (Georgiev, Lane, Rachuri, & Mascolo, 2014) have already shown that in a continuous audio sensing application, a DSP co-processor utilization can extend the battery life by 3 to 7 times. Similarly, GPU (Graphics Processing Unit) needs to be utilized to develop energy-optimized real-time applications for mobile phones (Cheng & Wang, 2011). A case study of GPU-assisted face feature extraction (Cheng & Wang, 2011) demonstrated a 3.98x reduction in total energy consumption. Smart watch and smart glass are relatively new in the market compared to smartphone. Similar for smartphone, energy conservation techniques (Wei, Yang, Rana, Chou, & Hu, 2012; Wei et al., 2013) (for both hardware and software) need to be developed for these two components to ensure their longer and reliable operation.

**Conclusion**

Telehealth is potentially an effective means of better managing patients at home. We have presented a number of key hurdles that need to be overcome for wider adoption of this technology. We have proposed an ensemble sensing framework

– wHealth, which collects information about physical, physiological and behavioural states of a person and facilitates effective and real-time communication channel between patient and clinician using monitors/sensors embedded in every day devices, such as watch, spectacles and phone. This framework enables a completely seamless delivery of telehealth services and imposes less maintenance overhead on the patients. Due to the non-stigmatising design, customisability and user-friendly interactions, this proposed framework has the potential to stimulate the widespread adoption of the telehealth technology. Additionally, we discuss two key challenges intrinsic to the wHealth platform and discuss potential overcoming strategies.